\documentclass[dvips,12pt]{article}
\usepackage{a4wide}
\usepackage{epsfig}
\usepackage{amsmath}
\usepackage{latexsym}
\usepackage{multirow}

\newcommand{\dd}{\mbox{{\rm d}}}

\def\deg{\hbox{$^{\circ}$}}

\newcommand{\GF}{G_{\rm F}}

\def\gsim{\mathrel{\rlap{\raise 2.5pt \hbox{$>$}}\lower 2.5pt
\hbox{$\sim$}}}
\def\lsim{\mathrel{\rlap{\raise 2.5pt \hbox{$<$}}\lower 2.5pt
\hbox{$\sim$}}}

\newcommand{\diag}{{\rm diag}}

\def\Month{\ifcase\month\or
January\or February\or March\or April\or May\or June\or
July\or August\or September\or October\or November\or December\fi}
\catcode`@=11
\def\citer{\@ifnextchar [{\@tempswatrue\@citexr}{\@tempswafalse\@citexr[]}}

\def\@citexr[#1]#2{\if@filesw\immediate
  \write\@auxout{\string\citation{#2}}\fi
  \def\@citea{}\@cite{\@for\@citeb:=#2\do
    {\@citea\def\@citea{--\penalty\@m}\@ifundefined
       {b@\@citeb}{{\bf ?}\@warning
       {Citation `\@citeb' on page \thepage \space undefined}}%
\hbox{\csname b@\@citeb\endcsname}}}{#1}}
\catcode`@=12

\begin{document}
 \thispagestyle{empty}
 \begin{flushright}
 {CERN-TH/99-285} \\[2mm]
% {hep-ph/0006185}
\end{flushright}
 \vspace*{2.0cm}
 \begin{center}
 {\bf \Large
 Universal quark-lepton mixing and determination of neutrino masses}
 \end{center}
 \vspace{1cm}
 \begin{center}
 Per Osland$^{a,}$\footnote{Work supported in part by the Research
Council of Norway.}
 \ \ and \ \
 Tai Tsun Wu$^{b,}$\footnote{Work supported in part by the United
States Department of Energy under Grant No.\ DE-FG02-84ER40158.}
\\
 \vspace{1cm}
$^{a}${\em
 Department of Physics, University of Bergen, \\
      All\'{e}gaten 55, N-5007 Bergen, Norway}
\\
\vspace{.3cm}
$^{b}${\em
 Gordon McKay Laboratory, Harvard University, \\
 Cambridge, Massachusetts 02138, U.S.A.}
\\
%\vspace{.3cm}
{\em and} \\
{\em
 Theoretical Physics Division, CERN,
CH-1211 Geneva 23, Switzerland}
\\
\end{center}
 \hspace{3in}
 \begin{abstract}
If three right-handed neutrinos are added to the Standard Model, then,
for the three known generations, there are six quarks and six leptons.
It is then natural to assume that the symmetry considerations that have
been applied to the quark matrices are also valid for the lepton mass
matrices. Under this assumption, the solar and atmospheric neutrino data
can be used to determine the individual neutrino masses.
Three minima have been found, using the $\chi^2$ fit, and, from these
minima, it is determined that the mass of the lightest neutrino is
$1.3\times10^{-3}$~eV, that of the next heavier neutrino is 
$1.3\times10^{-2}$~eV, while the mass of the heaviest neutrino is
$3.4\times10^{-2}$, $5.8\times10^{-2}$ or $9.4\times10^{-2}$~eV.
 \end{abstract}

\vfill
\begin{flushleft}
 {CERN-TH/99-285} \\[2mm]
 {March 2000}
\end{flushleft}
\newpage

\setcounter{footnote}{0}

%\pagestyle{myheadings}
%\markboth{\Draft{Draft}}{\Draft{Draft}}
%\markboth{\Draft{Version}}{\Draft{Version}}

%%%%%%%%%%%%%%%%%%%%%%%%%%%%%%%%%%%%%%%%%%%%%%%%%%%%%%%%%%%%%%%%%%%%%%%%
\section{Introduction}
\setcounter{equation}{0}
%%%%%%%%%%%%%%%%%%%%%%%%%%%%%%%%%%%%%%%%%%%%%%%%%%%%%%%%%%%%%%%%%%%%%%%%

In the work of Lehmann, Newton, and Wu \cite{LNW}, 
the Kobayashi-Maskawa \cite{KM} matrix is expressed
in terms of the masses of the three generations of quarks:
\begin{equation}
\begin{pmatrix} u \\ d\end{pmatrix}
\begin{pmatrix} c \\ s\end{pmatrix}
\begin{pmatrix} t \\ b\end{pmatrix}.
\end{equation}
This is accomplished by introducing a new horizontal symmetry.
Some of the earlier attempts in this direction are given in
\cite{FrXi,Fukugita}.

Recent experiments at Super-Kamiokande \citer{Super-K-prl,Super-K-spectrum}
indicate the presence of neutrino oscillations, which would imply that the 
neutrinos are not all massless.  If it is accepted that the neutrinos 
are not massless, then it is most natural in the Standard Model \cite{GSW}
to introduce three 
right-handed neutrinos in addition to the three known left-handed ones.  
In this way, there are six quarks and six leptons.\footnote{As shown 
in \cite{BHP}, neutrino-oscillation experiments cannot distinguish
between massive Majorana and Dirac neutrinos.}
In this paper, the 
consequences of a universal quark-lepton mixing are studied.  In other 
words, the method of \cite{LNW} is used to express the lepton KM matrix 
and the neutrino mixing matrix
in terms of the masses of the three generations of leptons:
\begin{equation}
\begin{pmatrix} \nu_e   \\ e  \end{pmatrix}
\begin{pmatrix} \nu_\mu \\ \mu \end{pmatrix}
\begin{pmatrix} \nu_\tau \\ \tau \end{pmatrix}.
\end{equation}
Of course the masses of the three charged leptons are accurately
known, leaving as unknown parameters the masses of the three neutrinos.
Thus there are three parameters to be determined instead of seven,
the three masses plus the four in the lepton KM matrix.

     It is the purpose of this paper to use the data from solar 
neutrinos \cite{Cl,SAGE,GALLEX,Super-K-sun} and atmospheric neutrinos
\cite{Super-K-prl,Super-K-spectrum}
to determine the three neutrino masses separately, not 
only the differences of their squares.

%%%%%%%%%%%%%%%%%%%%%%%%%%%%%%%%%%%%%%%%%%%%%%%%%%%%%%%%%%%%%%%%%%%%%%%%
\section{The mixing matrix}
\setcounter{equation}{0}
%%%%%%%%%%%%%%%%%%%%%%%%%%%%%%%%%%%%%%%%%%%%%%%%%%%%%%%%%%%%%%%%%%%%%%%%
The quark mixing matrix proposed by Lehmann et al.\ \cite{LNW}
relates the mixing to the actual quark masses.
These mass matrices are each given by four parameters
$a$, $b$, $c$, and $d$, and, when applied to the leptons without
modification, take the form
\begin{eqnarray}
\label{Eq:md-nondiag}
M(\ell)&=&\begin{pmatrix}
0    & d(\ell) & 0 \\
d(\ell) & c(\ell) & b(\ell) \\
0    & b(\ell) & a(\ell)
\end{pmatrix}, \\[3mm]
\label{Eq:mu-nondiag}
M(\nu)&=&\begin{pmatrix}
0    & id(\nu) & 0 \\
-id(\nu) & c(\nu) & b(\nu) \\
0    & b(\nu) & a(\nu)
\end{pmatrix},
\end{eqnarray}
with
\begin{equation}
\label{Eq:b-vs-c}
b^2(\ell)=8c^2(\ell), \qquad b^2(\nu)=8c^2(\nu).
\end{equation}

The diagonalization of these mass matrices \cite{Rosen}
is achieved by
the orthogonal matrices $R(\ell)$ and $R(\nu)$, where
\begin{eqnarray}
\label{Eq:diagonal-d}
M(\ell)&=&R(\ell)M_{\rm diag}(\ell)R^{\rm T}(\ell), \\[3mm]
\label{Eq:diagonal-u}
M(\nu)&=&\diag(-i,1,1)R(\nu)M_{\rm diag}(\nu)R^{\rm T}(\nu)\diag(i,1,1).
\end{eqnarray}
If there is $CP$ violation in the lepton sector, then the imaginary
entries in (\ref{Eq:mu-nondiag}) and (\ref{Eq:diagonal-u}) are required.
If by any chance there is no $CP$ violation in the lepton sector,
then both $i$ and $-i$ there should be replaced by 1.

The diagonal mass matrices have the form
\begin{equation}
M_{\diag}=\begin{pmatrix}
\lambda_1 & 0         & 0 \\
0         & \lambda_2 & 0 \\
0         & 0         & \lambda_3
\end{pmatrix}
=\begin{pmatrix}
m_1 &    0 & 0 \\
0   & -m_2 & 0 \\
0   &    0 & m_3
\end{pmatrix},
\end{equation}
where $\lambda_2<0$ and
\begin{equation}
\label{Eq:mass-order}
m_1\le m_3.
\end{equation}
We shall here only be concerned with $M(\nu)$ and $R(\nu)$.

In dealing with quarks \cite{LNW}, the observed quark masses allow
a much stronger form for the inequality (\ref{Eq:mass-order}), namely
\begin{equation}
\label{Eq:mass-order-quarks}
m_1\le m_2 \le m_3.
\end{equation}
The lack of direct experimental data on the neutrino masses implies
that (\ref{Eq:mass-order}) can be used, but not (\ref{Eq:mass-order-quarks}).
The first task is therefore to determine the allowed region in the space
$(m_1,m_2,m_3)$, which must be between those permitted
by (\ref{Eq:mass-order}) and (\ref{Eq:mass-order-quarks}).

The parameters $a$, $b$, $c$, and $d$ are related to the masses by the
following conditions,
\begin{alignat}{2}
\label{Eq:m1m2m3}
a+c&=\hphantom{\hbox{$-$}}S_1&\mbox{}=m_3-m_2+m_1,\hbox to .83in{}\nonumber
\\
8c^2+d^2-ac&=-S_2&\mbox{}=m_3m_2-m_3m_1+m_2m_1,\hbox to .2in{} \nonumber \\
ad^2&=-S_3&\mbox{}=m_1m_2m_3.\hbox to 1.23in{}
\end{alignat}
The cubic equation for the parameter $a$ is then
\begin{equation}
\label{Eq:cubic-a}
9a^3-17S_1a^2+(8S_1^2+S_2)a-S_3=0.
\end{equation}
Any real cubic equation can have either one or three real solutions.
Where there is {\it one} real solution, that one is negative,
and thus unphysical, as is seen from (\ref{Eq:m1m2m3}).
Where there are {\it three} real solutions, one of them is negative,
while two are positive.
We shall refer to these two positive solutions as Solution 1 (larger $a$)
and Solution 2 (smaller $a$).

These considerations can be used to determine the allowed physical
region in the $(m_1/m_3, m_2/m_3)$ plane, as shown in Fig.~\ref{Fig:disc}.
This region is only slightly larger than the triangle given by
the inequality (\ref{Eq:mass-order-quarks}), with two additional
regions, one where $m_2>m_3$ and the other a very small one with
$m_1>m_2$.
%%%%%%%%%%%%%%%%%%%%%%%%%%%%%%%%%%%%%%%%%%%%%%%%%%%%%%%%%%%%%%%%%%%%%%
\begin{figure}[htb]
\refstepcounter{figure}
\label{Fig:disc}
\addtocounter{figure}{-1}
\begin{center}
\setlength{\unitlength}{1cm}
\begin{picture}(12,9.5)
\put(2.0,1.0)
{\mbox{\epsfysize=9.0cm\epsffile{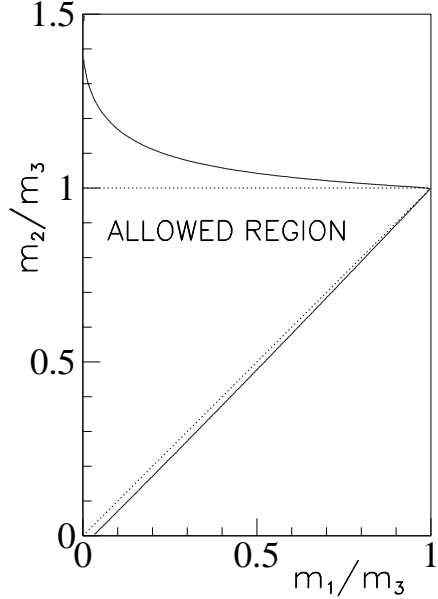}}}
\end{picture}
\vspace*{-8mm}
\caption{The allowed region for the three neutrino masses
$m_1$, $m_2$ and $m_3$ is within the {\it solid} contour.}
\end{center}
\end{figure}
%%%%%%%%%%%%%%%%%%%%%%%%%%%%%%%%%%%%%%%%%%%%%%%%%%%%%%%%%%%%%%%%%%%%%%

%%%%%%%%%%%%%%%%%%%%%%%%%%%%%%%%%%%%%%%%%%%%%%%%%%%%%%%%%%%%%%%%%%%%%%%%
\section{The three-family MSW mechanism}
\setcounter{equation}{0}
%%%%%%%%%%%%%%%%%%%%%%%%%%%%%%%%%%%%%%%%%%%%%%%%%%%%%%%%%%%%%%%%%%%%%%%%
The coupled equations satisfied by the three neutrino wave functions are
\cite{MSW}
\begin{equation}
\label{Eq:Schr-1}
i\,\frac{\dd}{\dd r}
\begin{pmatrix}
\phi_1(r) \\ \phi_2(r) \\ \phi_3(r)
\end{pmatrix}
=\left[
\begin{pmatrix}
D(r) & 0 & 0 \\
0 & 0 & 0 \\
0 & 0 & 0 \\
\end{pmatrix}
+\frac{1}{2p}
\begin{pmatrix}
M^2_{11} & M^2_{12} & M^2_{13}\\
M^2_{21} & M^2_{22} & M^2_{23}\\
M^2_{31} & M^2_{32} & M^2_{33}
\end{pmatrix}
\right]
\begin{pmatrix}
\phi_1(r) \\ \phi_2(r) \\ \phi_3(r)
\end{pmatrix},
\end{equation}
where
$D(r)=\sqrt{2}\,\GF N_e(r)$,
with $\GF$ the Fermi weak-interaction constant
and $N_e(r)$ the solar electron density at a distance $r$ from the center
of the sun.
Furthermore,
we denote $\nu_e=\phi_1$, $\nu_\mu=\phi_2$, $\nu_\tau=\phi_3$.

The evolution of the neutrino wave functions is determined by
the squared mass matrix,
\begin{equation}
[M(\nu)]^2
=\begin{pmatrix}
d^2 & cd & bd \\
cd & b^2+c^2+d^2 & b(a+c) \\
bd & b(a+c) & a^2+b^2
\end{pmatrix}
\equiv\begin{pmatrix}
M^2_{11} & M^2_{12} & M^2_{13}\\
M^2_{21} & M^2_{22} & M^2_{23}\\
M^2_{31} & M^2_{32} & M^2_{33}
\end{pmatrix},
\end{equation}
the neutrino momentum, $p$,
and the solar electron density.
Here, $M^2_{ij}\equiv(M^2)_{ij}$.
The eigenvalues of the squared mass matrix (multiplied by $r_0/2p$,
with $r_0$ defined below)
are denoted $\mu_1$, $\mu_2$, and $\mu_3$, and ordered such that
\begin{equation}
\mu_1 \le \mu_2 \le \mu_3.
\end{equation}

It is actually a good approximation to take an exponential electron
density, $N_e(r)=N_e(0)\exp(-r/r_0)$.
A fit to the solar density as given by \cite{BBP-98} leads to
$r_0=6.983\times10^4$~km.
For this case of an exponential solar density, the three-component
wave equation can be solved in terms of generalized
hypergeometric functions, ${}_2F_2$ \cite{OW-99}.

We scale and shift the radial variable, $u=r/r_0+u_0$,
with $u_0$ determined such that
\begin{equation}
D(0)r_0e^{u_0}=1.
\end{equation}
The above equation (\ref{Eq:Schr-1}) may then be written as
\begin{equation}
\label{Eq:Schr-2}
i\,\frac{\dd}{\dd u}
\begin{pmatrix}
\psi_1 \\ \psi_2 \\ \psi_3
\end{pmatrix}
=
\begin{pmatrix}
\omega_1+e^{-u} & \chi_2 & \chi_3 \\
\chi_2 & \omega_2 & 0 \\
\chi_3 & 0 & \omega_3 \\
\end{pmatrix}
\begin{pmatrix}
\psi_1 \\ \psi_2 \\ \psi_3
\end{pmatrix},
\end{equation}
where $\omega_2$ and $\omega_3$ denote the eigenvalues of the
$2\times2$ mass matrix involving $\psi_2$ and $\psi_3$.

With $x=e^{-u}$, one finds for
$\psi_1$ a differential equation of the form
\begin{equation}
\label{Eq:ode}
\left[\prod_{j=1}^3\left(x\frac{\dd}{\dd x}-i\mu_j\right)
-ix\prod_{j=2}^3\left(x\frac{\dd}{\dd x}-i\omega_j+1\right)\right]
\psi_1=0
\end{equation}
the solutions of which,
\begin{equation}
\psi_1(u)=\sum_{j=1}^3 C_j \psi_1^{(j)}(u),
\end{equation}
can be given in terms of generalized
hypergeometric functions ${}_2F_2$ as
\begin{equation}
\psi_1^{(j)}(u)=e^{-i\mu_ju}
{}_2F_2\biggl[
\begin{matrix}
1-i(\omega_2-\mu_j), & 1-i(\omega_3-\mu_j)\\
1-i(\mu_k-\mu_j),    & 1-i(\mu_\ell-\mu_j)
\end{matrix}
\bigg|
ie^{-u}
\biggr]
\end{equation}
with $k,\ell=1,2,3$, $k\ne j$, $\ell\ne j$.
For the other flavours, $\psi_2(u)$ and $\psi_3(u)$ are given by
similar expressions, with shifted parameters.

There is not much information about these functions ${}_2F_2$
in the literature \cite{Bateman}.
For the case of two flavours, the products in (\ref{Eq:ode}) go only up
to $j=2$, and a familiar confluent hypergeometric function ${}_1F_1$
(also denoted Whittaker function or parabolic cylinder function)
is obtained \cite{Petcov}.

In order to impose the boundary conditions that only {\it electron}
neutrinos are produced in the sun, we have to determine these functions
at large and negative values of $u$.
The series expansion is in principle convergent, but it is not practical
for large absolute values of both parameters and the argument.
Instead, methods have been developed for the evaluation by a combination
of relating them to ${}_3F_1$ and asymptotic methods \cite{OW-99}.

%%%%%%%%%%%%%%%%%%%%%%%%%%%%%%%%%%%%%%%%%%%%%%%%%%%%%%%%%%%%%%%%%%%%%%%%
\section{Results}
\setcounter{equation}{0}
%%%%%%%%%%%%%%%%%%%%%%%%%%%%%%%%%%%%%%%%%%%%%%%%%%%%%%%%%%%%%%%%%%%%%%%%
Since the only unknown parameters for the present theory of universal 
quark-lepton mixing are the three neutrino masses, it remains to 
determine these masses using experimental data on solar and atmospheric 
neutrinos.

For the atmospheric neutrinos we take the 8 data points
for $\nu_\mu$ and the 8 data points for $\nu_e$, as reported
recently \cite{Super-K-prl}.
These sixteen data points are treated as separate inputs, 
but we allow two overall normalization constants for the two sets 
of data.

For the solar-neutrino data, we use the total rates
from the Chlorine experiment \cite{Cl},
the Gallium experiments \cite{SAGE,GALLEX} (we average the two results)
and the Super-Kamiokande experiment \cite{Super-K-sun}.
We adopt the neutrino energy spectra and detector efficiencies
as given by Bahcall et al., and, for the latter detector,
we also include the neutral-current cross section \cite{Bahcall}.

The determination of the three neutrino masses is to be carried out 
by a $\chi^2$ fit to these pieces of data.  Since the number of degrees 
of freedom is 14, a good $\chi^2$ fit would give support to this idea 
that the mass mixing is universal for quarks and leptons.  For this 
purpose, we have scanned the entire ($m_1$, $m_2$, $m_3$) parameter 
space.  It is the necessity to cover this entire space that makes 
it essential to develop the theory described in Sec.~3 here.

It is found that there are three minima. They are:
\begin{alignat}{3}
&\text{(A) Solution 1:} &\qquad   
&\text{(B) Solution 2:}  &\qquad  
&\text{(C) Solution 2:} \nonumber \\
&m_1 = 0.0016~{\rm eV} &\qquad 
&m_1 = 0.0011~{\rm eV} &\qquad 
&m_1 = 0.0011~{\rm eV} \nonumber \\
&m_2 = 0.013~{\rm eV} &\qquad 
&m_2 = 0.013~{\rm eV} &\qquad 
&m_2 = 0.013~{\rm eV} \nonumber \\
&m_3 = 0.034~{\rm eV} &\qquad
&m_3 = 0.058~{\rm eV} &\qquad
&m_3 = 0.094~{\rm eV} \nonumber \\
&\text{with }\chi^2=31 &\qquad 
&\text{with }\chi^2=23 &\qquad 
&\text{with }\chi^2=20 \nonumber 
\end{alignat} 

A few simple conclusions can be drawn from this set of mass values.  
First, with 14 degrees of freedom, the three values of $\chi^2$ must 
be considered to be quite good. {\it This is evidence in favour of
universal quark-lepton mixing}.  In our mind the difference in these 
three values of $\chi^2$ is not significant.  Secondly,
the three values for $m_2$ are the same---a surprise to us.  
This presumably means that the relative accuracy in the present 
determination of $m_2$ is better than those of $m_1$ and $m_3$.  
This can be understood in the following way.  The relative
accuracy for $m_1$ is low because $m_1$ itself is quite small, 
and the value of $\chi^2$ is not very sensitive to such small masses.  
The average of these three $m_1$ is given in the abstract, and very 
roughly we guess the error to be about a factor of two.  
On the other hand, the uncertainly in the value of $m_3$ is simply due to 
the three minima giving different values.  It is a matter of which 
one of the three minima is physically the correct one.

For completeness, we give the rotation matrix $R(\nu)$.
The Euler angles $(\theta,\phi,\psi)$ \cite{Goldstein} are:
($22\deg,28\deg,-8\deg$) for (A), 
($35\deg,21\deg,-5\deg$) for (B), and
($31\deg,19\deg,-3\deg$) for (C).

To demonstrate the scanning of the entire region of allowed values 
of neutrino masses, we show in Fig.~\ref{Fig:Sol2-lvls-fin} an example 
of such a scan showing contours of constant $\chi^2$.  In this figure, 
the value of $m_3$ used is 0.058~eV, corresponding to the minimum 
(B) above.  It is from such scans that we know there are only three 
minima.  In this figure, some minor irregularities along the edges 
of the allowed region are artifacts of the finite grid spacing.
%%%%%%%%%%%%%%%%%%%%%%%%%%%%%%%%%%%%%%%%%%%%%%%%%%%%%%%%%%%%%%%%%%%%%%%%
\begin{figure}[htb]
\refstepcounter{figure}
\label{Fig:Sol2-lvls-fin}
\addtocounter{figure}{-1}
\begin{center}
\setlength{\unitlength}{1cm}
\begin{picture}(9.0,9.0)
\put(0.5,0.0){
\mbox{\epsfysize=9cm\epsffile{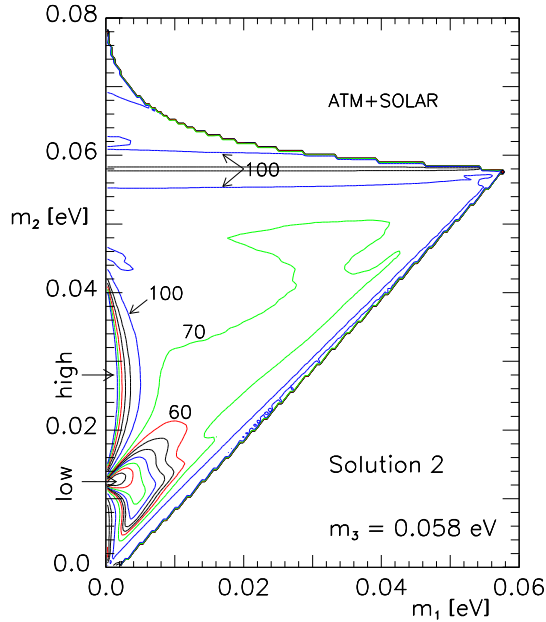}}}
\end{picture}
\hspace{5mm}
\caption{
Fits to the atmospheric- and solar-neutrino data.
Contours are shown at $\chi^2=30$, 35, \ldots, 60, 70, 100, 150, 
\ldots, 350.
(The outer contour outlines the boundary of the allowed region,
cf. Fig.~\ref{Fig:disc}.)}
\end{center}
\end{figure}
%%%%%%%%%%%%%%%%%%%%%%%%%%%%%%%%%%%%%%%%%%%%%%%%%%%%%%%%%%%%%%%%%%%%%%%%

It is difficult to compare the present result with the previously 
given allowed regions in the parameter space.
The reason is that the allowed regions have typically
been given on the basis of the mixing of two neutrino species,
while there is significant mixing among all three neutrinos,
or the studies concentrate on {\it either} solar {\it or} 
atmospheric neutrinos \cite{BKS,others}.
The following comparisons are nevertheless of interest.

(a) It is probably correct to compare the $m_3^2 - m_2^2$ here with 
the $\delta m^2$ from atmospheric-neutrino data.  In this case, the 
$m_3^2 - m_2^2$ for the three minima covers roughly the allowed range of 
$\delta m^2$, with minimum (B) near the center, and minima (A) and (C) 
near the edges of the allowed region.

(b) It is probably not too far wrong to identify $m_2^2 - m_1^2$ 
here with the $\delta m^2$ from solar-neutrino data.  If so, the values 
are reasonably close, except with the so-called 
``low-mass--low-probability'' (LOW) solution \cite{BKS}.

(c) It is more difficult to discuss the strength of the coupling.  
It can nevertheless be concluded, in the context of solar neutrinos, 
that the present solutions are significantly closer to the 
``large-mixing-angle'' (LMA) solution than to 
the ``small-mixing-angle'' (SMA) solution.

In connection with the problem of distinguishing between the 
three minima, more accurate data from Super-Kamiokande and 
related experiments are needed.  
Also, a better understanding of the high-energy $hep$ neutrino flux 
\cite{BKS} could make the electron recoil energy spectrum useful
for this purpose. Furthermore, important 
information may be forthcoming also from the long-baseline 
experiments \cite{LBLE} and exotic atom experiments \cite{exotic-atoms}.
\bigskip

\leftline{\bf Acknowledgments}
\par\noindent
We are greatly indebted to Professor Harry Lehmann and 
Professor Jack Steinberger for the most helpful discussions.  
We would also like to thank Dr.\ Steve Armstrong, 
Professor Alvaro de Rujula, Dr.\ Conrad Newton,
Professor Gabriele Veneziano, 
and especially Geir Vigdel, for useful discussions.  
One of us (TTW) wishes to thank 
the Theory Division of CERN for its kind hospitality.
\bigskip

%%%%%%%%%%%%%%%%%%%%%%%%%%%%%%%%%%%%%%%%%%%%%%%%%%%%%%%%%%%%%%%%%%%%%%%%
%\clearpage

\end{document}